\documentstyle[epsf,bibnorm]{lamuphys}
\makeatletter
\let\chapter\hid@chapter
\makeatother
\begin{document}
\pagenumbering{arabic}

\title{Missing Momentum Distributions and Nuclear Transparency in Semi-inclusive $A(e,e'p)X$ Processes
\protect\footnote{Presented by C. Ciofi degli Atti at the Elba Workshop on Electron-Nucleus Scattering, EIPC June 22-26 1998}
}

\author{Claudio\,Ciofi degli Atti\inst{1}, Hiko\, Morita\inst{1,2} and Daniele\, Treleani\inst{3}}

\institute{ Department of Physics, University of Perugia, and Istituto Nazionale di Fisica Nucleare, Sezione di Perugia, Via A. Pascoli, I--06100 Perugia, Italy
\and
Sapporo Gakuin University, Bunkyo--dai 11, Ebetsu 069, Hokkaido, Japan
\and
Department of Theoretical Physics, University of Trieste, Strada
Costiera 11, and Istituto Nazionale di Fisica Nucleare, Sezione di Trieste, I--34014, Trieste,Italy}

\titlerunning{Missing momentum distributions}

\maketitle 

\begin{abstract}
 A linked cluster expansion for the distorted one-body mixed density matrix is obtained within the Glauber multiple scattering theory with correlated wave functions. The nuclear transparency for $^{16}O$ is calculated using realistic central and non-central correlations. The convergence of the expansion is investigated in the case of $^4He$ for which the transparency and the distorted momentum distributions are calculated to all order in the correlations using a variational wave function obtained from realistic NN interactions. The important role played by non central correlations is illustrated.
\end{abstract}

\section{Introduction}

Many calculations aimed at investigating the  role played by ground state nucleon--nucleon correlations on the nuclear transparency and distorted momentum distributions, have  been so far performed within different approaches (see e.g. \cite {fan:nikolayev}, \cite {fan:bentra}, \cite {fan:rinat}, \cite {fan:seki}). It is the aim of this contribution to present a novel approach based upon a linked-cluster expansion of the one body distorted mixed density matrix starting from realistic correlated wave functions and Glauber multiple scattering operators. Our paper is organised as follows: in Section 1 the basic elements of the theory are presented;  the formal developments of the  linked-cluster  expansion are illustrated in Section 2; the results of the calculations of the nuclear transparency for $^{16}O$, using the lowest order cluster expansion, are presented in Section 3;   the results of calculations of  the nuclear transparency and the distorted momentum distribution for $^4He$, taking correlations into account to all orders using a variational realistic wave function,  are presented in Section 4; the Summary and Conclusions are given in Section 5.

\section{The missing momentum distributions and nuclear transparency within a linked cluster expansion for Glauber correlated wave functions}

The basic quantity appearing in the definition of the distorted momentum distribution is the distorted one-body mixed density matrix
\begin{eqnarray}
\rho_D (\vec r,\vec r')= \frac {<\Psi \bf S^+ \hat{O}(\vec r,\vec r') {\bf S'}\Psi'>}{<\Psi\Psi>}
  \label{eq:rodi}
   \end{eqnarray}
where $\Psi$ is the nuclear wave function, $\bf S$ is the operator which introduces final state interacions (FSI), and $\hat{O}(\vec r,\vec r')$ is the one-body mixed density operator
\begin{eqnarray}
\hat{O}(\vec r,\vec r')= \sum_i\delta(\vec r_i -\vec r)\delta(\vec r_i' -\vec r')\prod_{j\not= i}\delta(\vec r_j -\vec r_j')
\label{eq:rodiop}
   \end{eqnarray}
In Eq. \ref{eq:rodi} and in the rest of the paper, the primed quantities have to be evaluated at $\vec r'$. The operator ${\bf S}$ will be evaluated within Glauber multiple scattering theory, i.e.

\begin{eqnarray}
{\bf S}=\prod_{j>1}^A G(1,j)  \qquad G(1,j)=1+ \Gamma (1,j)\Theta(1-j)
  \label{eq:essegi}
   \end{eqnarray}
where $\Gamma (1,j)$ is the Glauber profile and $1$ refers to the hit nucleon. By integrating $n_D(\vec p)$
\begin{eqnarray}
n_D(\vec p)={(2 \pi)^{-3/2}} \int e^{i \vec p(\vec r -\vec r')}\rho_D (\vec r,\vec r') d\vec r d\vec r'
  \label{eq:ennedi}
   \end{eqnarray}

one obtaines the nuclear transparency $T$ defined as follows

\begin{eqnarray}
\int n_D(\vec p) d\vec p = \int \rho_D (\vec r,\vec r') d\vec r d\vec r' 
\int e^{i \vec p(\vec r -\vec r')}d\vec p = \int \rho_D (\vec r)d \vec r
  \label{eq:intennedi}
   \end{eqnarray}
i.e.

\begin{eqnarray}
T = \frac{
\int\rho_D (\vec r)d \vec r}{A} = 1+ \Delta T
  \label{eq:ti}
   \end{eqnarray}
where $\Delta T$ originates from FSI. Note that when the FSI are absent, the usual nucleon momentum distribution $n(\vec p)$ is recovered, with normalisation

\begin{eqnarray}
\int n(\vec p) d\vec p = \int \rho (\vec r)d \vec r = A
  \label{eq:norma}
   \end{eqnarray}

We  have evaluated (\ref{eq:rodi}) using for $S$ the form ( \ref{eq:essegi})
and for the nuclear wave function the following form

\begin{eqnarray}
\Psi = {\rm {\hat{S}}} \Big [ \prod_{i<j} f ^n (r_{ij})\hat {O}^n (ij)\Big ] 
 \Psi_o
  \label{eq:psi}
   \end{eqnarray}
where $\rm {\hat{S}}$ is the symmetrisation operator, $\Psi_o$   the Slater determinant describing the nucleon independent particle motion, and $f ^n (r_{ij})$ the correlation function associated to the operator $\hat{O}^n (ij)$; if $\hat{O}^n (ij) = 0$ for $n>1$, the usual Jastrow wave function is recovered.
 We have developed a linked cluster expansion in the  quantity
$\eta (iji'j') = f(ij)f(i'j') - 1$ which includes at each order of the correlation the Glauber operator to all orders, and the  result at first order in $\eta (iji'j')$ reads as follows

\begin{eqnarray}
\rho_D (1,1') &\simeq &<\Psi_o \mid \prod G^+G { \hat O }(1,1')\mid \Psi_o>\nonumber\\ &+&
 <\Psi_o \mid \prod G^+G \sum \eta{ \hat O }(1,1')\mid \Psi_o>\nonumber\\ &-&
<\Psi_o \mid \sum\eta\mid\Psi_o><\Psi_o\mid\prod G^+G \mid \Psi_o>
  \label{eq:rounounop}
   \end{eqnarray}
Note that the above expansion is linked, in that the unlinked contributions in the second and third terms 
cancel exactly. Eq. \ref{eq:rounounop}
allows a diagrammatic representation, where the basic elements are the following ones

\begin{figure}
\centerline{
\epsfysize=10cm \epsfbox{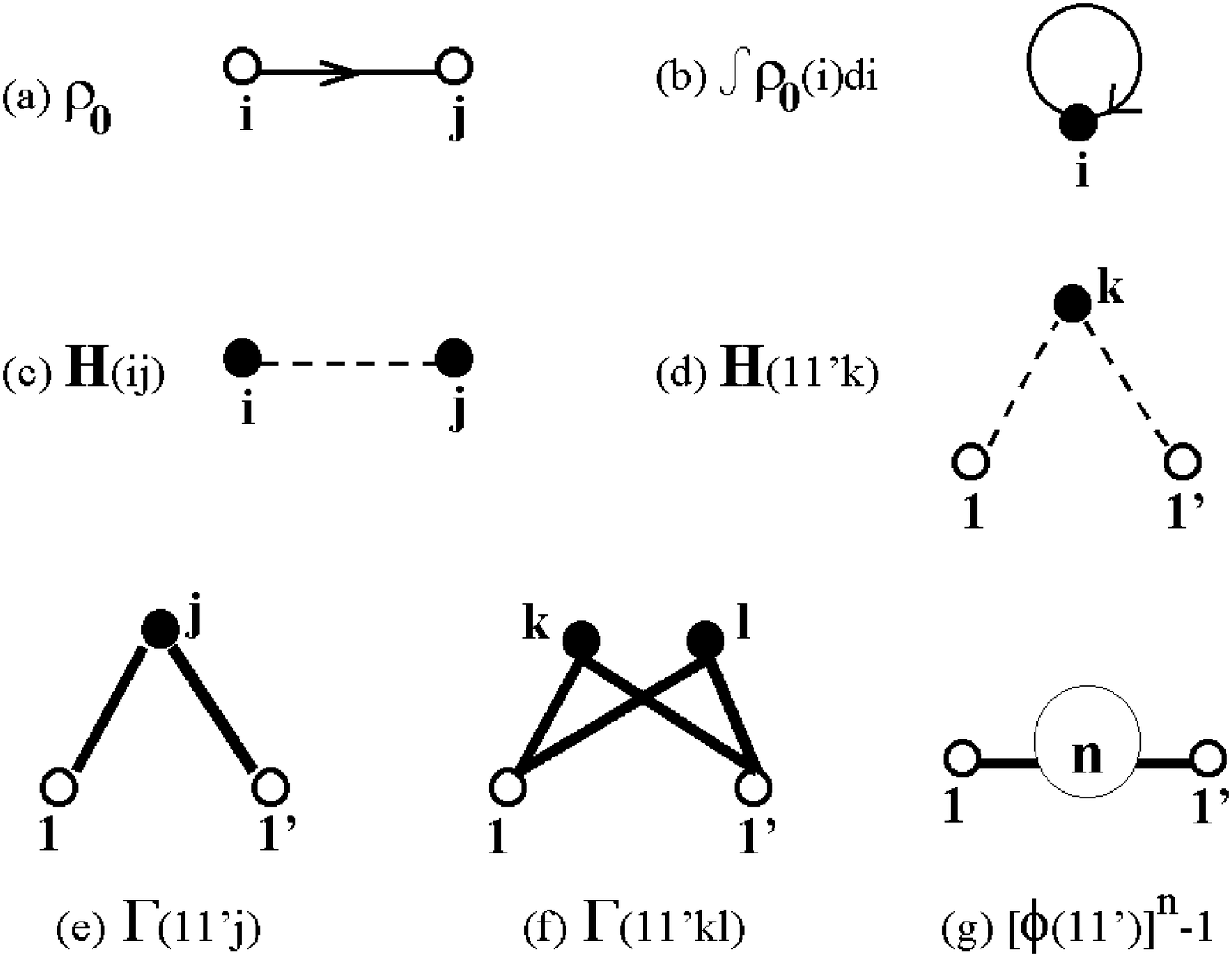}  
}
\vspace{2 cm}
\caption[ ]{ The various diagrams corresponding to the terms in(\ref{eq:elementi}).  }
\label{diagrammi}
\end{figure}
\begin{eqnarray}
&(a)& \qquad\rho_o (i,j)= 4\sum_ {\alpha \in F}\phi_{\alpha}^*(\vec r_i)\phi_{\alpha}(\vec r_j)\nonumber\\
&(b)& \qquad\int \rho_o(i)di\nonumber\\
&(c)& \qquad{\bf H}(ij)\nonumber\\
&(d)& \qquad{\bf H}(11'k)\label{eq:elementi}\\
&(e)& \qquad{\bf \Gamma}(11'j)={\bf \Gamma}(1j)+{\bf \Gamma}(1'j)+{\bf \Gamma}(1j){\bf \Gamma}(1'j)\nonumber\\ 
&(f)& \qquad{\bf \Gamma}(11'kl)=\sum{\bf \Gamma} +\sum{\bf \Gamma}{\bf \Gamma}+...\nonumber\\
&(g)& \qquad\big [\Phi (11')\big]^n\equiv\Big [\int \rho_o(j) G^+(1j)G(1'j)dj\Big]^n\nonumber
  \end{eqnarray}
where $\rho_o (i,j)$ is the shell model one-body density matrix and ${\bf H}(ij)$ is the correlation term. The diagrammatic representation of the above quantities is presented in Fig. \ref{diagrammi}.
\begin{figure}
\centerline{
\epsfysize=10cm \epsfbox{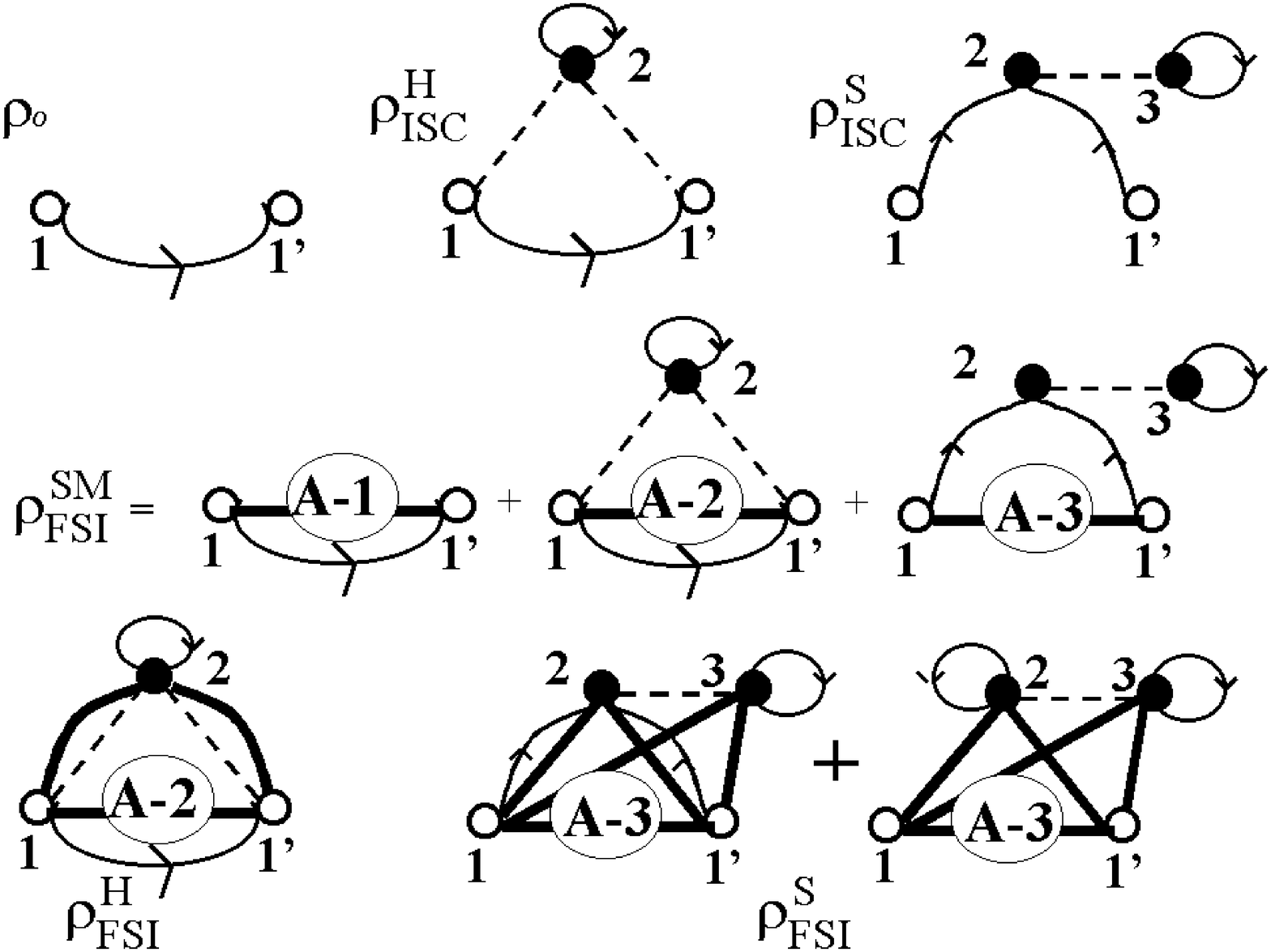}  
}
\vspace{2 cm}
\caption[ ]{ The various diagrams corresponding to the terms in(\ref{eq:rounounod}). }
\label{diagrammiro}
\end{figure}

\begin{figure}
\centerline{
\epsfysize=10cm \epsfbox{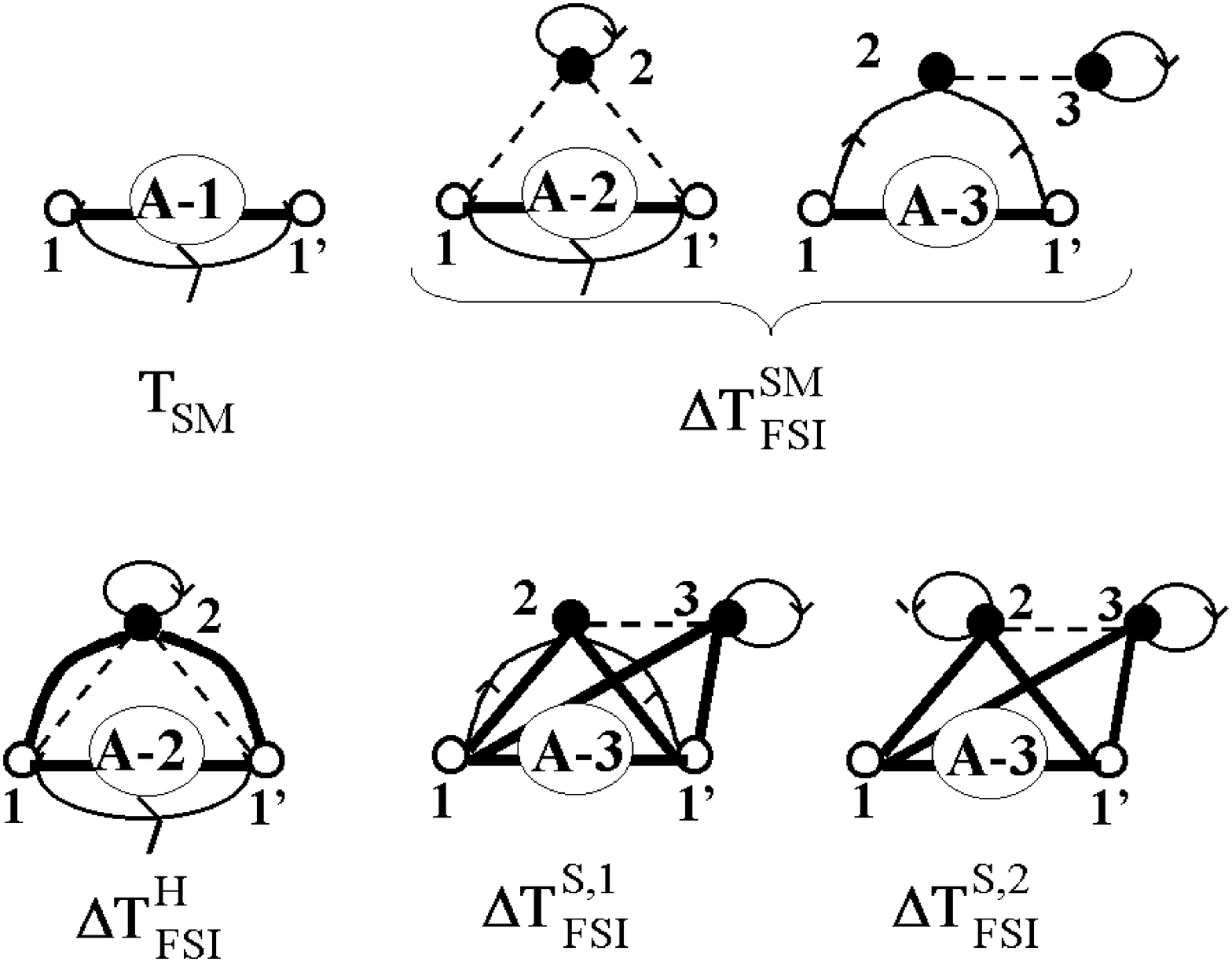}  
}
\vspace{2 cm}
\caption[ ]{ The various diagrams corresponding to the terms of the nuclear transparency (\ref{eq:trasp}). }
\label{diagrammiti}
\end{figure}

The distorted one-body mixed density matrix assumes the following form

\begin{eqnarray}
\rho_D (1,1') \simeq \rho_o +\rho_{ISC}^H +\rho_{ISC}^S +\rho_{FSI}^{SM} +\rho_{FSI}^H  +\rho_{FSI}^S 
\label{eq:rounounod}
   \end{eqnarray}
where $H$ and $S$ stand for {\it hole} and {\it spectator }, respectively.
In Eq. \ref {eq:rounounod} we have separated the contributions from initial state correlations (ISC) from the ones in which the final state interaction (FSI) acts between shell model (SM) uncorrelated nucleons ($\rho_{FSI}^{SM}$) and the ones in which FSI acts between nucleon 1, correlated with nucleon 2,  and (A-2) uncorrelated nucleons ($\rho_{FSI}^H$) and between (A-3) uncorrelated nucleons and nucleon 1 with nucleon 2, correlated with nucleon 3 ($\rho_{FSI}^S$). The diagrammatic representation of the various terms of (Eq. \ref{eq:rounounod}) are presented in  Fig. \ref{diagrammiro}. When the FSI is set equal to zero ($G=1$ or $\Gamma=0$), the last three terms of  (\ref{eq:rounounod}) disappear and the first three terms will coincide with the ones considered  in several papers (see e.g. \cite{fan:benhar}\cite{fan:co})to calculate  the correlated momentum distributions. The details of the derivation of (Eq. \ref{eq:rounounod}) will be given elsewhere \cite{fan:clada} (note that  the diagrammatic representation in  Fig. \ref{diagrammiro} is strictly
valid for large A, when $(A-3) \approx (A-2)  \approx (A-1) \approx A$).

\section{The nuclear transparency for $^{16}O$}

The nuclear transparency has been calculated by Eq. (\ref{eq:ti}). Note that since the linked cluster expansion we are using is a number conserving one, the terms $\rho_{ISC}^H$ and   $\rho_{ISC}^S$ give  equal and opposite contributions to the integral in Eq. (\ref{eq:ti}), so that $\Delta T$ gets contribution only from the terms $\rho_{FSI}^{SM}$, $\rho_{FSI}^H$, and   $\rho_{FSI}^S$; correspondingly, the nuclear transparency can be represented in the following way 

\begin{eqnarray}
T= 1
+ \Delta {T^{SM}_{FSI}} + \Delta {T^{H}_{FSI}} + \Delta {T^{S,1}_{FSI}} +\Delta {T^{S,2}_{FSI}},
  \label{eq:trasp}
   \end{eqnarray}
where we have separated the spectator contribution in two parts which, as will be seen later on , cancel to a large extent. Let us reiterate that $\Delta {T^{SM}_{FSI}}$ includes Glauber FSI to all order between the hit nucleon and uncorrelated nucleons.
The diagrammatic representation of Eq. (\ref{eq:trasp}) is given in  
Fig. \ref{diagrammiti}. Calculations have been performed by parametrising the Glauber profile as follows

\begin{eqnarray}
\Gamma (b) = \frac{\sigma_{tot}(1-i\alpha)}{4\pi b_o^2}e^{- b^2/(2b_o^2)}
  \label{eq:gamma}
   \end{eqnarray}
with $\sigma_{tot} =43 mb$, $\alpha = -0.33$ and $b_o = 0.5 fm$. As far as the nuclear wave function is  concerned, we have considered two different cases:
$i)$ the phenomenological Jastrow wave function with central correlations with frequently used in the calculations of the transparency (see e.g. \cite{fan:nikolayev}); $ii)$ the wave function (Eq. \ref{eq:psi}) corresponding to the Reid V8 interaction \cite{fan:vuotto},  with single particle and correlation parameters determined from the minimisation of the nuclear hamiltonian \cite{fan:benhar}.  The results of the calculations are presented in Table \ref{table1}.

\begin{table}
\caption[dummu6]{The nuclear transparency for $^{16}O$ (Eq. \ref{eq:trasp}).} 
\begin{flushleft}
\renewcommand{\arraystretch}{1.2}
\begin{tabular}{llllllllllll}  
\hline\noalign{\smallskip}
   &   \qquad  $T_{SM}   \qquad \qquad  $  &  $\Delta {T^{SM}_{FSI}}$ \qquad\qquad &   $\Delta {T^{H}_{FSI}}$ \qquad\qquad &  $\Delta{T^{S,1}_{FSI}}$ \qquad\qquad & 
$\Delta{T^{S,2}_{FSI}}$ \qquad\qquad& $T$  \qquad\qquad \\
\hline
Central    & \qquad 0.51 & 0.020 & 0.032  & --0.013 & 0.022 & 0.57 \\
Realistic  & \qquad 0.51 & 0.003  & 0.009 & 0.001 & --0.001 & 0.52 \\
\noalign{\smallskip}\hline
\end{tabular}
\renewcommand{\arraystretch}{1}
\label{table1}
\end{flushleft}
\end{table} 

The results of calculations, which are presented in Table \ref{table1}, deserve the following comments:
\begin{enumerate}
\item Within the phenomenological central correlation approach, the effects of correlations on the nuclear transparency  is sizeable (about 12\%)
\item The contribution of the spectator term  is almost zero, originating from two terms of opposite sign, and the effect of FSI  within correlated nucleons is almost entirely due to the hole contribution
\item Non central correlations affect very sharply the nuclear transparency, in that the overall effect of correlations reduces to about 2\%, with the hole contribution remaining the dominant one and the spectator contribution cancelling out.
\end{enumerate}
\begin{figure}
\centerline{
\epsfysize=10cm \epsfbox{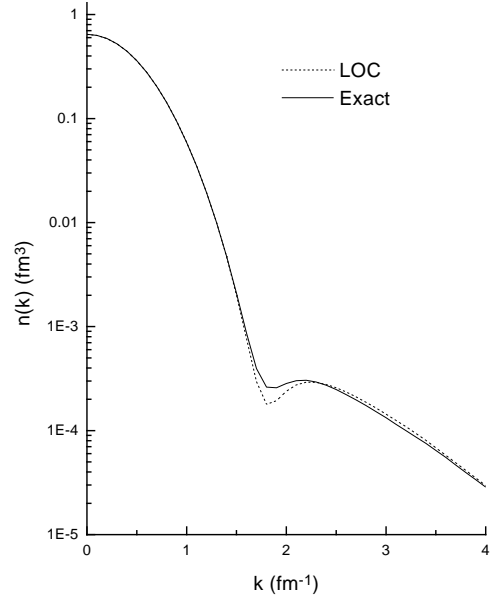}  
}
\vspace{0.5 cm}
\caption[ ]{ The undistorted momentum distribution $n(\vec k)$ }
\label{ennek}
\end{figure}

\begin{figure}
\centerline{
\epsfysize=10cm \epsfbox{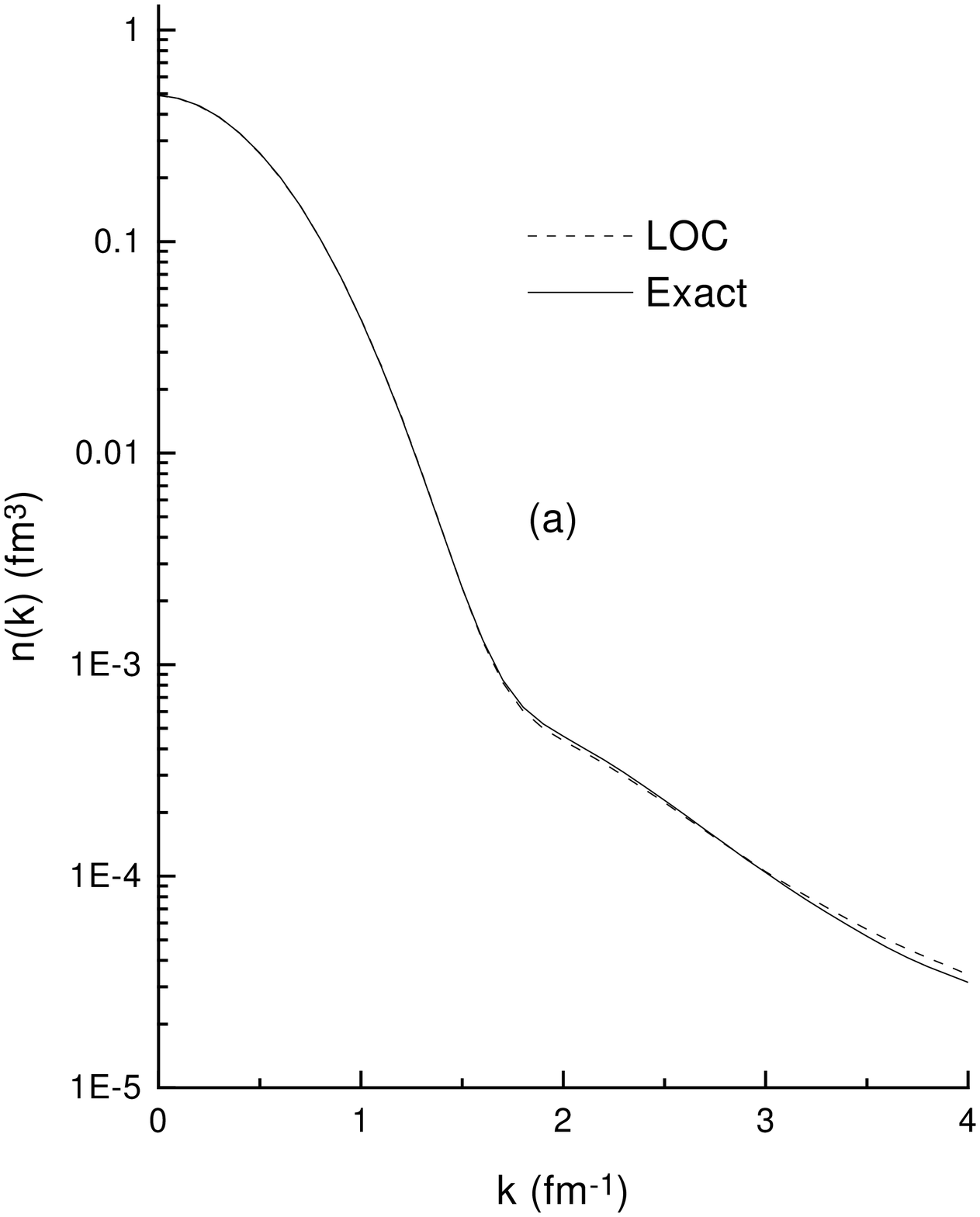}  
\epsfysize=10cm \epsfbox{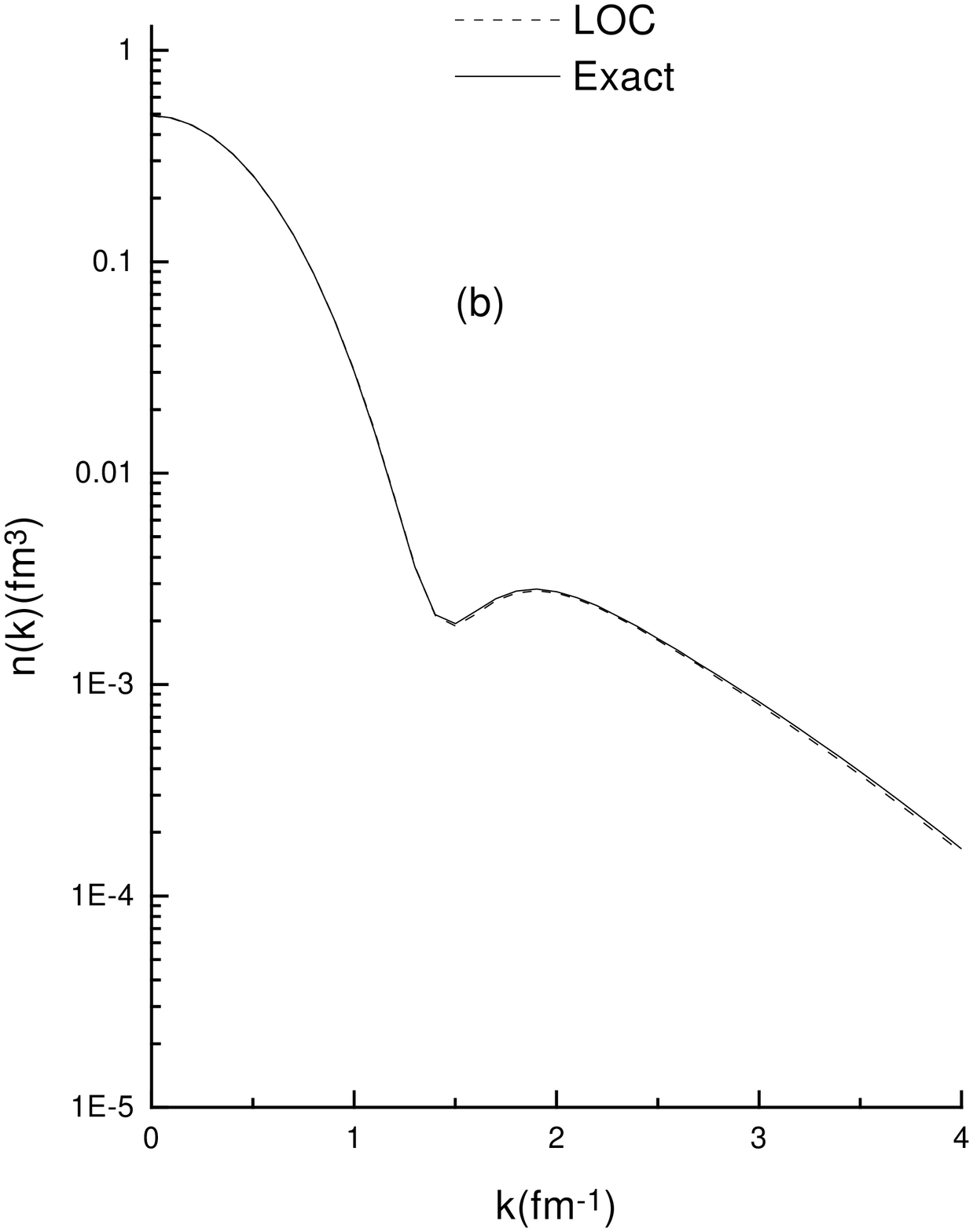} 
}
\vspace{1 cm}
\caption[ ]{ The distorted  longitudinal $(\theta =0^o)$ (a) and perpendicular $(\theta =90^o)$ (b) momentum distributions calculated within the Jastrow approach: the dashed curve (LOC) refers to the lowest order cluster expansion and the full curve (Exact) to the complete result }
\label{ennefsi_J}
\end{figure}

\begin{figure}
\centerline{
\epsfysize=10cm \epsfbox{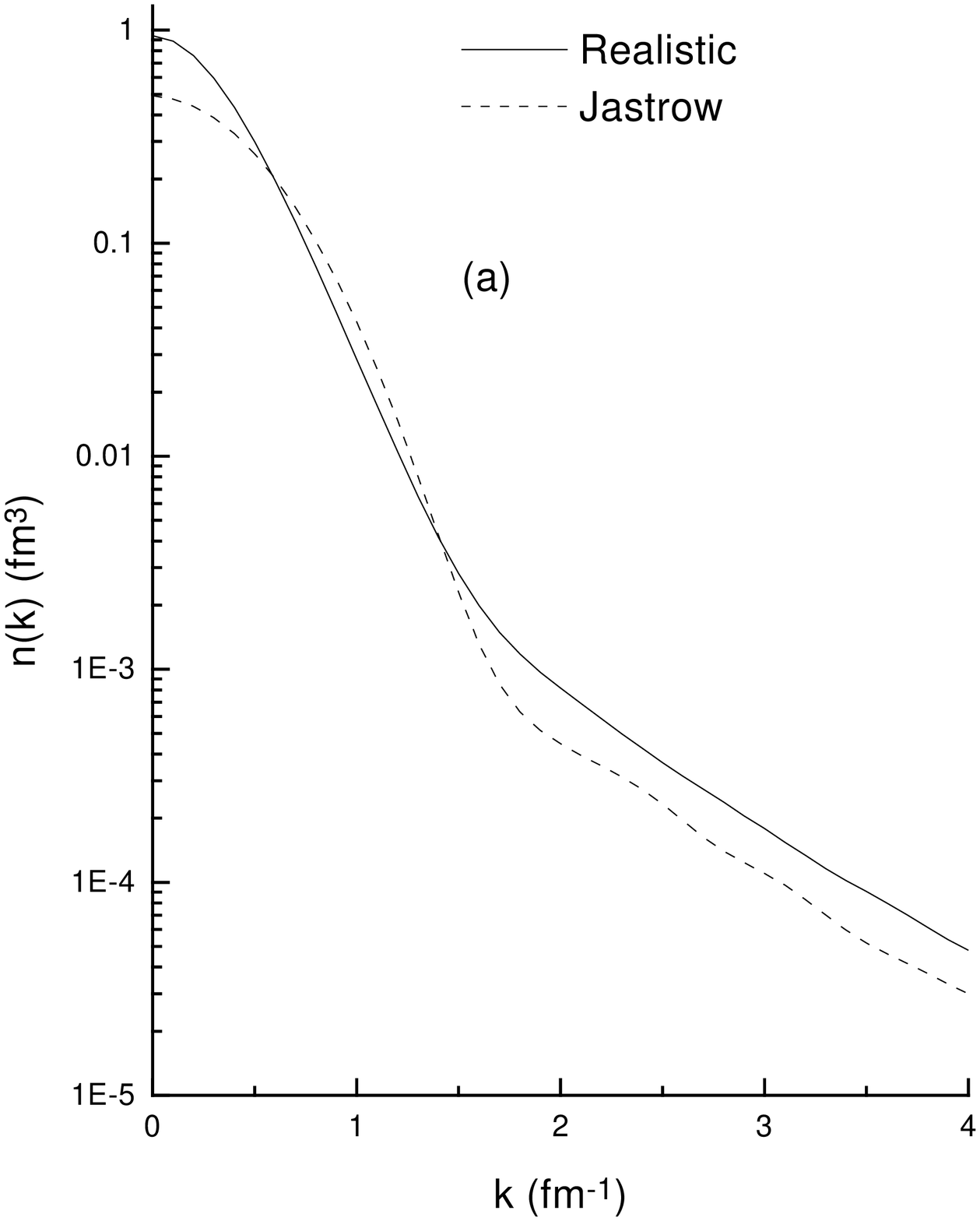}  
\epsfysize=10cm \epsfbox{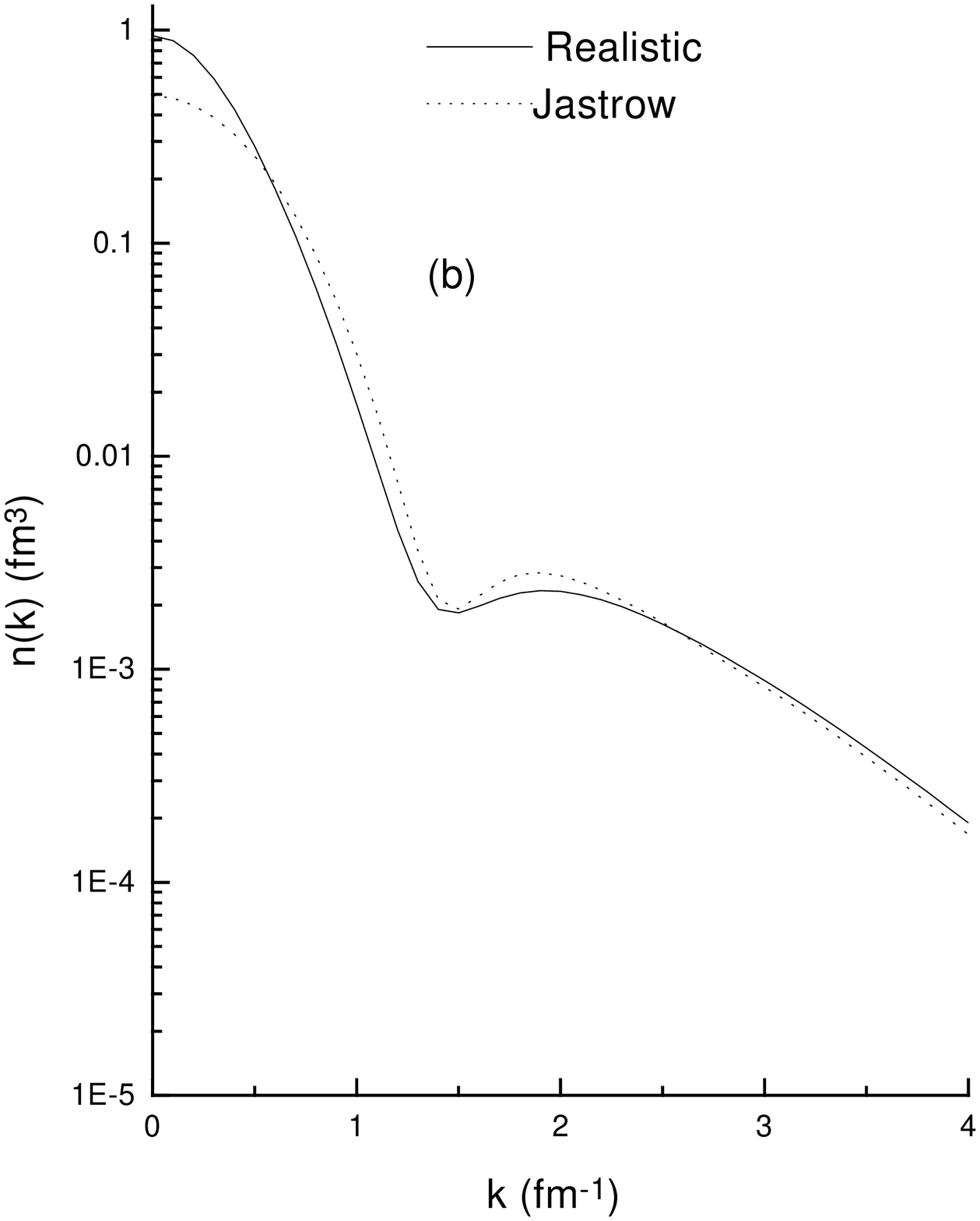} 
}
\vspace{1 cm}
\caption[ ]{ The distorted  longitudinal $(\theta =0^o)$ (a) and perpendicular $(\theta =90^o)$ (b) momentum distributions calculated within the Jastrow (dashed) and  realistic (full) approaches }
\label{ennefsi_JR}
\end{figure}

\begin{figure}
\centerline{
\epsfysize=10cm \epsfbox{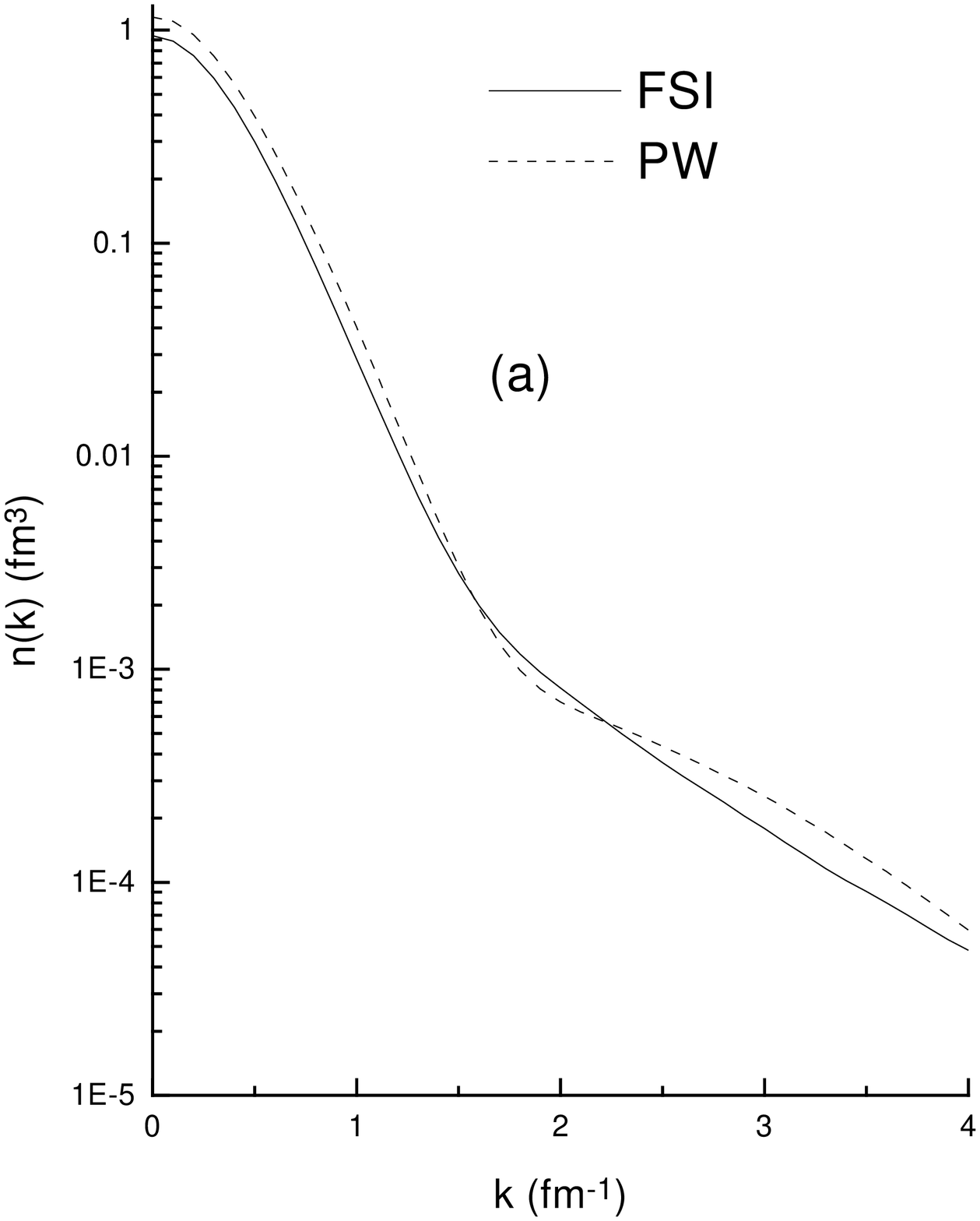} 
\epsfysize=10cm \epsfbox{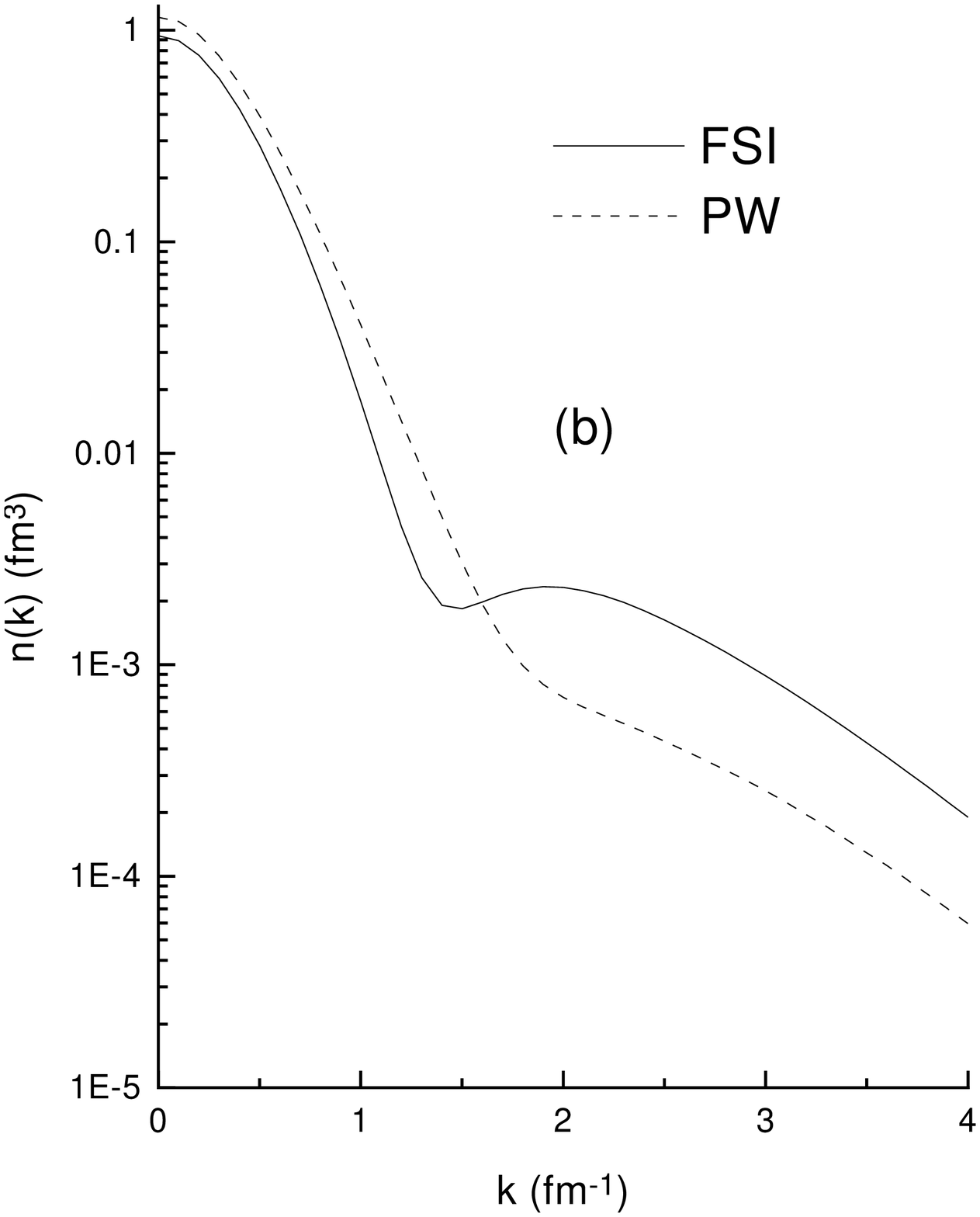}   
}
\vspace{1 cm}
\caption[ ]{ The distorted  longitudinal $(\theta =0^o)$ (a) and perpendicular $(\theta =90^o)$ (b) momentum distributions (full) compared with the undistorted ones (dotted). Calculations correspond to the realistic wave function}
\label{n-pw-fsi_R}
\end{figure}

\begin{figure}
\centerline{
\epsfysize=10cm \epsfbox{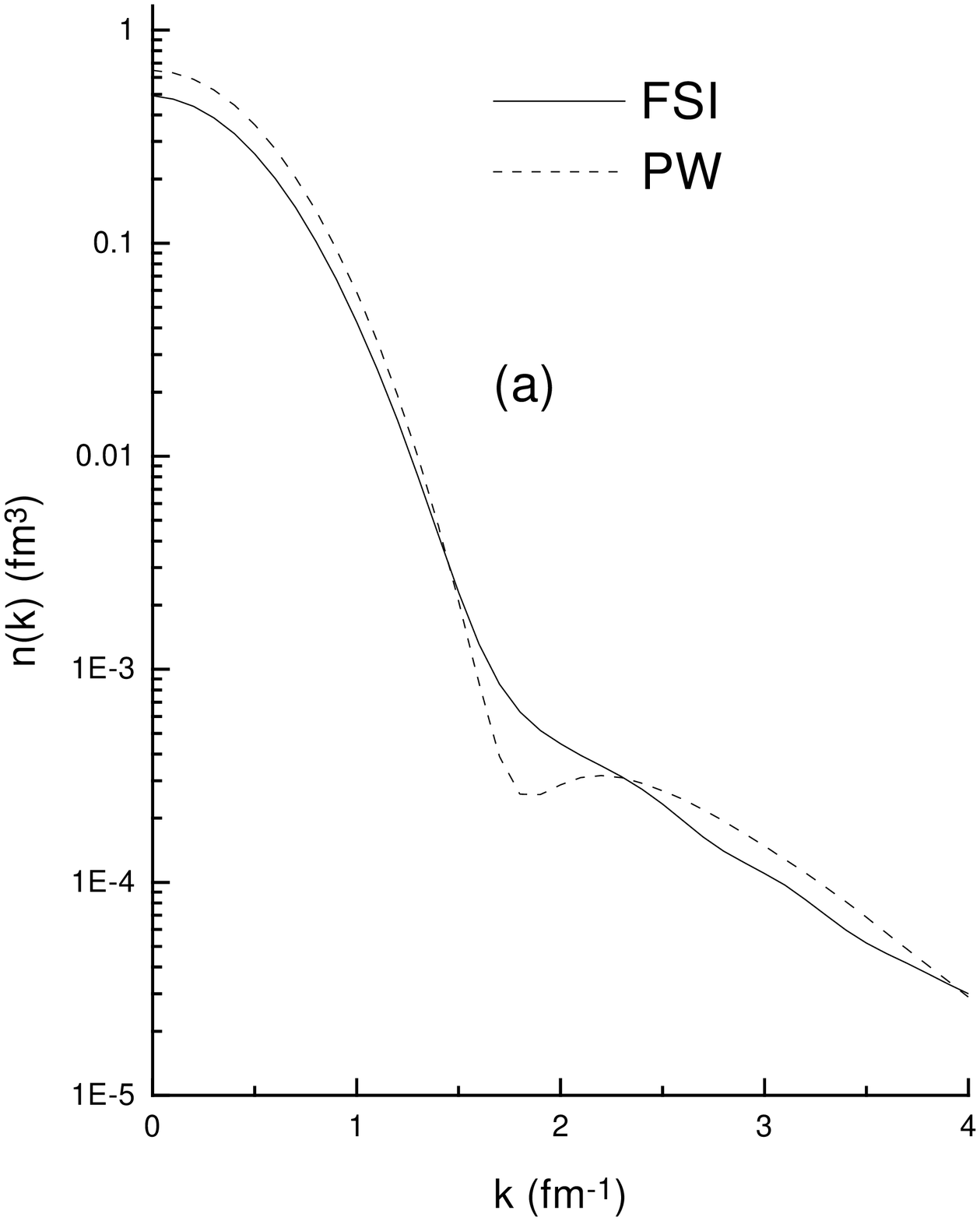}  
\epsfysize=10cm \epsfbox{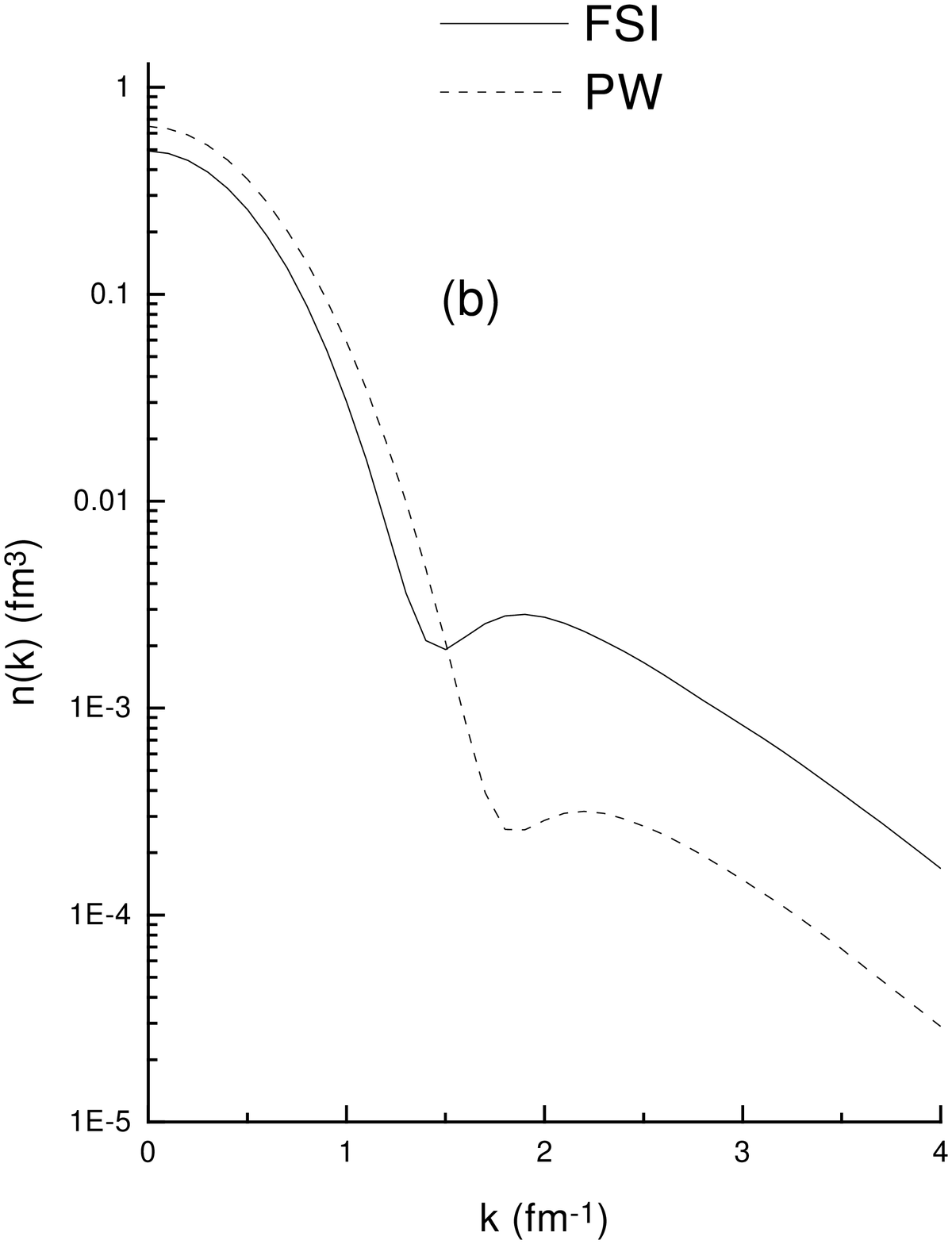}  
}
\vspace{1 cm}
\caption[ ]{ The same as Fig.7 but with the Jastrow wave function containing only central correlations}
\label{n-pw-fsi_J}
\end{figure}
\begin{figure}
\centerline{
\epsfysize=10cm \epsfbox{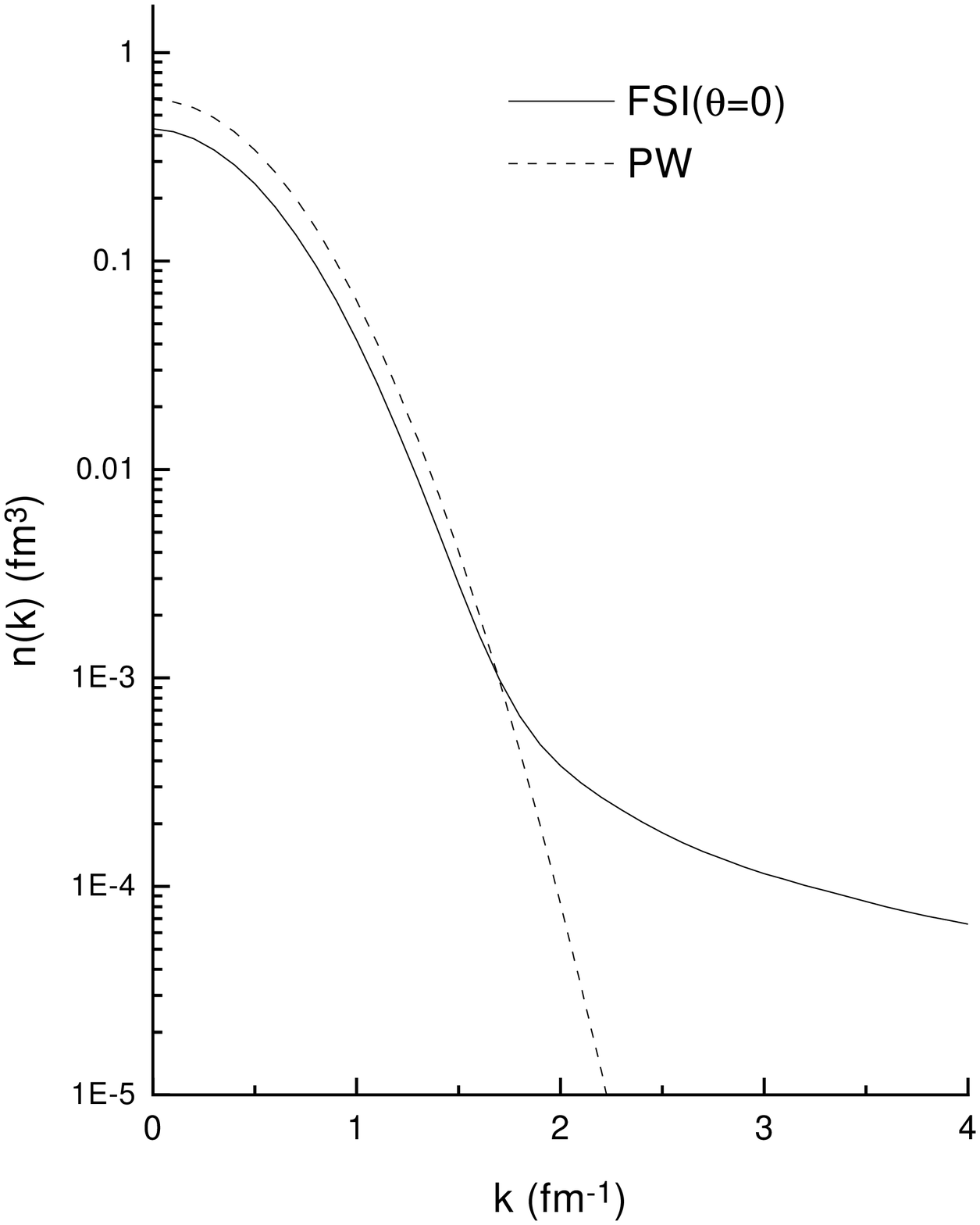}  
}
\vspace{1 cm}
\caption[ ]{ The distorted  longitudinal $(\theta =0^o)$ compared with the undistorted ones (dotted). Calculations correspond to harmonic oscillator wave functions}
\label{n-pw-fsi_HO}
\end{figure}

The above results are at variance with the conclusions reached in \cite{fan:nikolayev}, where a phenomenological central correlation approach, not based upon a linked-cluster expansion, has been used. In fact in \cite{fan:nikolayev} a cancellation between hole and spectator contributions has been claimed to occur, giving as a result, a negligible contribution of correlations to the transparency.

Although previous calculations of various observables \cite{fan:co} speaks in favour of a good convergence of the number-conserving cluster expansion, the effect of the higher order terms on the nuclear transparency is a topic worth being investigated. To this end, in the next Section the results for the nuclear transparency and distorted momentum distributions of  $^4He$, obtained exactly with
 a correlated wave function of the type (\ref{eq:psi}), will be presented.

\section{The nuclear transparency and distorted momentum distributions for $^{4}He$}

In the case of $^4He$ we have calculated \cite{fan:cmt1} the momentum distributions and  the nuclear transparency by evaluating (Eq. \ref{eq:rodi}) within two different approaches: $i)$the central Jastrow approach within  both the  lowest order cluster expansion, and
the exact evaluation of $\rho_D$; $ii)$ the  ATMS method \cite{fan:atms} which provides a realistic wave function in the form 
\begin{eqnarray}
\Psi_{ATMS} = \sum_{ij} \omega (ij) \prod_{kl \not= ij} u(kl)
  \label{eq:ATMS}
   \end{eqnarray}
where $\omega (ij)$ and $u(kl)$ are state-dependent functions.
 The results for the nuclear transparency are presented in Table 2.
\begin{table}
\caption[dummu6]{The nuclear transparency for $^4He$ corresponding to the lowest order and  the full  Jastrow calculation, and to  the realistic ATMS wave function corresponding to the Reid V8 interaction \cite{fan:vuotto}} 
\begin{flushleft}
\renewcommand{\arraystretch}{1.2}
\begin{tabular}{llllllllllllllll}  
\hline\noalign{\smallskip}
   &   \qquad \qquad \qquad \qquad   $T_{SM}   \qquad \qquad  $  &  $\Delta {T^{LOC}_{FSI}}$ \qquad\qquad &   $\ {T^{LOC}_{FSI} }$ \qquad\qquad & 
 $T^{EX}$  \qquad\qquad \\
\hline
Jastrow        & \qquad \qquad \qquad \qquad 0.75 & 0.046 & 0.80 &  0.80      \\
Realistic  & \qquad \qquad \qquad \qquad 0.75 & 0.017 &0.77 & 0.78 \\
\noalign{\smallskip}\hline
\end{tabular}
\renewcommand{\arraystretch}{1}
\label{table2}
\end{flushleft}
\end{table} 

The following comments are in order:
\begin{enumerate}
\item The lowest order cluster expansion reproduces almost completely the exact result.
\item The value of the nuclear transparency sensitively depends upon the nature of  correlations, in that, when realistic non central correlations are also considered, the transparency is strongly reduced, becoming, as in the case of $^{16}O$, of the order of 3\%.
\item The convergence of the cluster expansion is a very good one, both in the Jastrow and the realistic cases.
\end{enumerate}

We have also calculated the distorted momentum distributions within the above three approaches. i. e.:
\begin{enumerate}
\item The cluster expansion with central correlation.
\item The full Jastrow approach with central correlations.
\item The realistic calculation with the ATMS wave function.
\end{enumerate}

The distorted momentum distributions can be written in the form $n_D (\vec p)=
n_D (\vec p_{\perp},p_{\parallel})$; we have calculated the {\it parallel} 
 $(\theta =0^o, p_{\perp}= 0)$, {\it antiparallel}  $(\theta =180^o, p_{\perp}= 0)$,
and {\it perpendicular}$   (\theta =90^o, p_{\parallel}= 0)$ kinematics, where  $\theta$ is the angle between the three-momentum transfer $\vec q$ and $\vec p$. The validity of the cluster expansion in the calculation of the momentum distribution $n(\vec p)$ is shown in Fig. \ref{ennek}.
It can be seen that the lowest order cluster expansion works extremely well. This is not the case if a non converging cluster expansion is used, e.g. the one in which the numerator and the denominator are expanded independently \cite{fan:nikolayev}. We have worked out the distorted momentum distribution by keeping into account  FSI exactly and by using both the lowest order cluster expansion and the complete calculation.
The results are presented in Fig. \ref{ennefsi_J} where the longitudinal and perpendicular momentum distributions are shown.

It can again be seen that the lowest order cluster expansion works extremely well even when FSI are present.

In
 Fig. \ref{ennefsi_JR} the phenomenological distorted momentum distributions are compared with the realistic ones. It turns out that realistic non central correlations have sizeable effects on $n_D (\vec k)$, in particular on the high momentum components.
Let us now discuss in more details the effects of FSI on the distorted momentum distributions. To this end, in Fig. \ref{n-pw-fsi_R} the undistorted and distorted longitudinal and perpendicular momentum distributions corresponding to the realistic wave functions, are compared;  the effect of FSI on the former is very small, whereas on the latter is, on the contrary, very large.
 The same comparison is shown for the Jastrow wave functions in Fig. \ref{n-pw-fsi_J}, and it can be seen that FSI appreciably depends upon the ground state wave function which is used to calculate $n_D$:  when realistic ground state correlations are considered, the effect of FSI is strongly reduced. Such a behaviour could  in principle be foreseen , for  the main effect of FSI is to produce high missing momentum components, so that, if these are already present in the ground state,  FSI should produce a relatively lower  increase of high momentum components. Such a feature is clearly illustrated in Fig. \ref{n-pw-fsi_HO}
,  where one can see that if high momentum components are totally absent in the ground state (which, of course, represents a very  unrealistic case), the high missing momentum components are entirely created by FSI. It can therefore be concluded that the investigation of the relevance of FSI in semi-inclusive $A(e,e'p)X$ processes, based upon unrealistic ground state wave functions, can lead to unreliable conclusions.

\section{Summary and Conclusions}

Our work can be summarised as follows:
\begin{enumerate}
\item A   linked cluster expansion has been developed which includes both initial state correlations and final state interactions.
\item Calculations show that  the spectator contribution almost cancels out and the main effect of correlations is due to the hole contribution.
\item Whereas the effect of central correlations on the transparency is to increase it by about 10\%, when realistic central and non central correlations are considered, the effect reduces to about 3\%. Such a result holds both for $^{16}O$, treated within the cluster expansion, and for $^4He$,treated within an exact approach.
\item The comparison between the cluster expansion and the  exact calculations shows a very good agreement, indicating a rapid convergence of the number conserving cluster expansion we have worked out.
\item In the case of $^4He$, the effect of FSI is greatly reduced when realistic 
wave functions are used and becomes a relatively small correction  for the longitudinal momentum distributions while it remains important for the 
perpendicular ones. The effect of FSI on the distorted momentum distributions for complex nuclei will be presented elsewhere.
\end{enumerate}



\begin{thebibliography}{99}
%
\bibitem{fan:nikolayev} 
N. N. Nikolayev et al, Phys. Lett. {\bf B317},281 (1993) 
%
\bibitem{fan:bentra}
O. Benhar et al, Phys. Rev. Lett. {\bf 69},881 (1992)
\bibitem{fan:rinat}
A. S. Rinat and B. K. Jennings, Nucl. Phys. {\bf A568},873 (1994)
%
\bibitem{fan:seki}
R. Seki et al, Phys.Lett.  {\bf B383},133 (1996)
%
\bibitem{fan:benhar} 
O. Benhar, C. Ciofi degli Atti, S. Liuti and G. Salme, Phys.  Lett. {\bf 28}, 885 (1986)
%
\bibitem{fan:co} 
G. C\'o et al 
 Phys. Rev.{\bf 69}, 981 (1997)
%
\bibitem{fan:clada}
C. Ciofi degli Atti and D. Treleani, to appear
%
\bibitem{fan:vuotto}
I.E. Lagaris and V.R. Pandharipande, Nucl. Phys. {\bf A359}, 331 (1981);
O. Benhar et al, Phys. Lett. {\bf B359},8 (1995)
%
\bibitem{fan:cmt1}
C. Ciofi degli Atti, H. Morita and D. Treleani, to appear
%
\bibitem{fan:atms}
M. Sakai, I. Shimodaya, Y. Akaishi, J. Hiura and H. Tanaka, Prog. Theor. Phys.
Suppl. No. {\bf 56}, 32 (1974)\\
H. Morita, Y. Akaishi, O. Endo and H. Tanaka, Prog. Theor. Phys. {\bf 78}, 1117 (1987) 
%
\end{thebibliography}
\end{document}